\documentclass[aps,prc,showpacs,a4paper,twocolumn,dvipdfmx,nofootinbib]{revtex4}
\usepackage[dvipdfmx]{graphicx}
\usepackage{amsmath}
\usepackage{amsfonts}

\begin{document}

\title{Symmetry projection to coupled-cluster singles and doubles wave function through the Monte Carlo method}

\author{Takahiro Mizusaki$^1$ and Peter Schuck$^{2,3}$}

\affiliation{
 $^1$Institute of Natural Sciences, Senshu University, Tokyo 101-8425, Japan\\
 $^2$Universit\'e Paris-Saclay, CNRS-IN2P3, IJCLab, 91405 Orsay cedex, France\\
 $^3$Univ. Grenoble Alpes, CNRS, LPMMC, 38000 Grenoble, France}

\begin{abstract}
A new method for calculating the symmetry-projected energy of coupled-cluster singles and doubles (CCSD) wave function through the Monte Carlo method is proposed. We present benchmark calculations in considering the three-level Lipkin model which is a simple and minimal model with two phases: spherical and deformed ones. It is demonstrated that this new method gives good ground state energy and low-lying spectra.
\end{abstract}


\maketitle

\section{Introduction}
To solve many-body problems, the use of intricate trial wave function is absolutely desirable, while computability limits its functional form. To enlarge the power of trial wave functions has been one of the central issues of many-body physics. Let us look back at some well-established theories.
In the Raleigh-Ritz variational approach, Hartree-Fock (HF) and Hartree-Fock-Bogoliubov (HFB) wave functions correspond to  mean-field theories and have been well-developed \cite{RS80} in the past. 
Beyond mean-field, symmetry restoration is essential and is well compatible with HF and HFB approaches. Symmetry projection has also been extensively studied \cite{RS80,SD19}. 
On the other hand, as a non-variational approach, the coupled-cluster method  \cite{FC58,CK60,KLZ78,HN78} is quite prominent and has been well developed with a broad field of applications. 
Its trial wave function is a mean-field wave function multiplied by an exponential with many-body correlations, $e^Z$, which has an elaborate
structure.
The coupled-cluster method, however, seems not to be well suited for symmetry projection.

In the present study, we unite these two approaches.
We start with the coupled-cluster singles and doubles (CCSD) wave function, 
and optimize it by the coupled-cluster method.
Then, we apply the symmetry projection operator to it 
and evaluate the expectation value of the Hamiltonian.
Of course, due to the $e^Z$ correlation factor, the straightforward utilization inevitably leads to huge and mostly intractable computational difficulties. Therefore, we introduce the Monte Carlo approach.  
We propose a new integration of coupled-cluster wave functions with symmetry projection via the Monte Carlo method.  

To evaluate this new method, we use the three-level Lipkin model \cite{HY74,HN78,HB00}, which is a generalization of the Lipkin-Meshkov-Glick model \cite{LMG65}.
This model is simple but exhibits a spherical-deformed phase transition.  
For the case of degenerate single-particle energies of the upper two levels, 
there is an exchange symmetry, which can be handled by the symmetry projection method. Thus the three-level Lipkin model is a minimal model for our purpose. 
With numerical investigations, we will test the performance of  our new method. 

We also mention that recently the use of symmetry projection in the two-level Lipkin model has been studied in \cite{WS17} and symmetry projections have been implemented in the coupled-cluster method in \cite{QH18,QHDS19,BD21}. 
However, our approach is quite different from those studies. 

In the symmetry projection, we use spherical and deformed bases in a mixed way.
For the spherical basis, the creation and annihilation operators are denoted by $c_i^\dag$ and $c_i$, while for deformed basis, they are denoted by $a_i^\dag$ and $a_i$. 
The deformed operators are canonically related to the spherical operators as $a_i=\sum_j D_{ij} c_j$.
The true vacuum is represented as $|-\rangle$. 
Hereafter, a state defined by spherical operators is denoted as 
$|i_1,\cdots,i_m\rangle=c_{i_1}^\dag\cdots c_{i_m}^\dag|-\rangle$, while a state defined by deformed operators is denoted as  $|k_1,\cdots,k_m)=a_{k_1}^\dag\cdots a_{k_m}^\dag|-\rangle$.

\section{Method}
\subsection{CCSD and CCD wave functions}
We begin to consider the coupled-cluster (CC)  method \cite{FC58,CK60,KLZ78,HN78}.
Based on a HF wave function $|\psi_0 )$, we define the CC wave function 
$|\psi_{cc})$ with the many-body $Z$ operator as
$
  |\psi_{cc} )=e^{Z}|\psi_0 ).
$
Z is defined as
 $\sum \chi_{p_1,p_2,\cdots h_1,h_2,\cdots} a_{p_1}^\dag a_{p_2}^\dag\cdots a_{h_1} a_{h_2} \cdots$, where $p$'s ($h$'s) stand for unoccupied (occupied) orbits. 
The summation is arranged as  $Z= Z_1+Z_2+\cdots$, where the $Z_k$ are defined as  $Z_k=\sum_{i}x_i^{(k)}Z_i^{(k)}$ and the
$Z_i^{(k)}$ are the different $k$-body operators and  $x_i^{(k)}$ are their parameters.
We limit $Z$ by up to two-body terms ($k=2)$ due to the increasing numerical complexity in the applications for higher values of $k$, which we call the CCSD approximation. 

The parameters $x$ are determined in the following way. 
If $ |\psi_{cc})$ is an exact wave function, we have 
\begin{equation}
  H|\psi_{cc})=Ee^{Z}|\psi_0),
\end{equation}
where $E$ is the exact energy and $H$ is the Hamiltonian expressed with the deformed operators.
The equation corresponds to a non-Hermitian-type eigenvalue problem as follows  
\begin{equation}
   \bar{H}|\psi_0)=E|\psi_0),
   \label{hbareq}
\end{equation}
where the transformed Hamiltonian $\bar{H}=e^{-Z}He^{Z}$ is non-Hermitian.
The energy is obtained by 
\begin{equation}
 ( \psi_0 |\bar{H}|\psi_0)=E,
\label{cc_energy}
\end{equation} 
where we refer to it as the CC energy hereafter.
Multiplying  Eq.(\ref{hbareq}) from the left with  $( \psi^*| $   , the coefficients $x$ are determined by   
\begin{equation}
  (\psi^* |\bar{H}|\psi_0)=0,
  \label{cceq}
\end{equation}
where $( \psi^*| $ is any state orthogonal to $ ( \psi_0| $.
Within the CCSD approximation, all operators $Z_1^{(k)}$, $Z_2^{(k)}$ and their parameters  $x_1^{(k)}$, $x_2^{(k)}$ are renamed as
 $z_i (i=1,\cdots n_0)$ and $x_i (i=1,\cdots n_0)$
where $n_0$ is the number of parameters. The mean-field wave function $|\psi_0)$ is simply shown by $|0)$ as $0$-th basis state. 
We evaluate the following matrix elements for an excited state  $( k|$  
 as $a_k=( k| H|0)$, $ b_{ki}=( k| [H,z_i]|0)$ and
$c_{kij}=( k|[ [H,z_i],z_j]|0)$
where $i,j,k=1,\cdots n_0$.
Equation (\ref{cceq}) is then rewritten as 
\begin{equation}
  a_k +\sum_i b_{ki}  x_i 
  +\frac{1}{2} \sum_{ij}  c_{kij} x_i =0.
\end{equation} 
The parameters $x$ can be obtained iteratively and we, thus, can evaluate the CC energy with the CCSD wave function. 

We add the following iterative procedure to eliminate the one-body operator $Z_1$ by changing the mean-field for later convenience.  
The exponential of the $Z_1$ operator changes the mean-field determinant $|0)$ into another mean-field one $|0')= e^{Z_1}| 0)$, that is, the Slater determinant
$D$ is modified by $e^{Z_1}$ and is changed into $D'$.
Then we recalculate all the matrix elements with this new $D'$ and solve the coupled-cluster equations. This procedure is repeated iteratively.
After convergence,  we obtain a new mean-field $D$ and the corresponding CCSD wave function as $e^{Z_2}|0)$ where  $|0)$ stands for the optimal mean-field state.
We refer to this form as the CCD wave function, which is the starting point of this study.

\subsection{Symmetry projection of CC wave function}

In the coupled-cluster method, the energy is given by Eq.(\ref{cc_energy}).
This method, however, does not satisfy the Raleigh-Ritz variational principle. Therefore, we evaluate the energy directly as,
\begin{equation}
E= \frac{(\psi | H|\psi )}{(\psi |\psi )},
\label{var_energy} 
\end{equation}
where $|\psi)$ stands for the CCSD or CCD wave function.
This energy gives an upper limit to the exact energy, unlike the coupled-cluster method. Hereafter we refer to it as the RR energy 
to distinguish it from the CC energy. 
 
Next, we introduce the symmetry projection operator $P^L$, which is a projection onto a state with a good quantum number  $L$. For example, in nuclear structure physics, angular momentum projection has been often utilized.
For a continuous symmetry, the symmetry projection operator is generally given by 
\begin{equation}
P^L=\frac{1}{\mathcal{N}}\int d\mu\mathcal{W}(L,\mu )R(\mu) ,
\end{equation}
where $\mathcal{N}$ is a normalization,  $\mathcal{W}$ is a weight function
and $R(\mu)$ is given by $e^{i\hat{O}\cdot\mu}$ where $\hat{O}$ is generally defined in terms of 
spherical operators, not by deformed ones.
The integration is carried out over the parametrization $\mu$ of the continuous group. 
We will show an example below in section III.
If $|\psi)$ is a superposition of several states with different $L$,
the projection operator $P^{L}$ extracts $|\psi,L)$ with a definite quantum number $L$  as
$|\psi,L)=P^{L}|\psi)=\frac{1}{\mathcal{N}}\int d\mu \mathcal{W}(L,\mu )R(\mu)|\psi)$.

The RR energy $E_{L}$ with the projected coupled-cluster wave function is given by
\begin{equation}
E_{L}=\frac{( \psi |H P^{L}|\psi ) }{( \psi|P^{L} |\psi )},
\label{projected_energy}
\end{equation}
where we refer to it as projected energy. If we follow the terminology 
of projection method e.g., variation-after-projection (VAP) or 
projection-after-variation (PAV),
we call this procedure projection-after-coupled-cluster (PACC). 
Note that this evaluation is usually carried out through spherical operators although the above equation is expressed in the deformed basis.
To evaluate the projected energy, we need the matrix elements with the projected wave function as
\begin{equation}
( \psi |
\left[
\begin{matrix}
1 \\
H
\end{matrix}
 \right] 
P^{L}|\psi)=\frac{1}{\mathcal{N}}\int d\mu \mathcal{W}(L,\mu)
( \psi|
\left[
\begin{matrix}
1 \\
H
\end{matrix}
 \right]
| \psi(\mu ) ).
\label{def_proj_mat}
\end{equation}
where the rotated state $|\psi( \mu) \rangle$ is defined by
\begin{equation}
 |\psi( \mu ) ) = e^{i\hat O\mu }|\psi ).
 \label{rotated_wf}
\end{equation}
This is the standard way for  projection calculations. In realistic applications with huge Hilbert spaces, this projection is generally not feasible because $|\psi)$ includes $e^{Z_2}$ in the CCD wave function, and evaluation of deformed operators through spherical operators additionally increases computational efforts.

\subsection{Monte Carlo Procedure}

To carry out the symmetry projection, we introduce the Monte Carlo procedure.  Hereafter we choose the label $s$ to specify each basis of the spherical representation. 
Similarly, we choose the label  $d$ to specify each basis in the deformed mean-field representation.  
The coupled-cluster wave function is given as $|\psi)$ with the deformed operators.

Inserting the unity operator $\sum_s |s\rangle \langle s|=1$, the projected energy in Eq.(\ref{projected_energy}) can be rewritten as 
\begin{equation}
E_L=\sum_{s}\rho_L(s){\cal E}_L(s).
\label{def_mc_energy}
\end{equation}
For the spherical basis $s$, we define the projected local energy ${\cal E}_L(s)$  as 
\begin{equation}
{\cal E}_L(s)=\sum_{s'} h_{s,s'} \frac{ \langle s'|P^L|\psi ) }
{\langle s|P^L|\psi )},
\label{localenergy}
\end{equation}
where $ h_{s,s'}$ is the Hamiltonian matrix element in the spherical representation and is generally very sparse.
We can also define the projected density $\rho_L(s)$ as
\begin{equation}
\rho_L(s)=\frac{|\langle s|P^L|\psi )|^{2}}
{\sum_{s}|\langle s|P^L|\psi)|^{2}},
\label{rho}
\end{equation}
where $\rho_L(s)\geq 0$ and $\sum_{s}\rho_L(s)=1$.
This property allows us to stochastically generate the distribution of $s$ according to Eq.(\ref{rho}) by the Markov chain Monte Carlo (MCMC) method. 
Note that this technique has been presented
in  cases with pair-condensates \cite{BM06,TI08,VMC12,VMC18}.
Therefore, by applying the Monte Carlo sampling,
the symmetry projected energy can be estimated as
\begin{equation}
E_L\sim \frac{1}{N_0}\sum_i {\cal E}_L(s_{i}),
\label{mc_energy}
\end{equation}
where  $N_0$ is a number of Monte Carlo samples and we refer to it as the MC energy hereafter.

Next, we investigate the projected overlap $\langle s|P^L|\phi)$ in detail. By inserting $\sum_d |d)(d|=1$, the projected overlap is rewritten as 
\begin{equation}
 \langle s| P^{L}|\psi)=\sum_d \langle s| P^{L}|d)(d |\psi).
 \label{sPLpsi}
\end{equation}
The correlation part of the overlap, $(d |\psi)=(d|e^{Z_2}|0)$ is easily computed because $Z_2$ is also represented with the deformed operators and $P^{L}$ does not operate on $|\psi)$.
The symmetry projection is easily carried out in the form $\langle s|P^L|d)$, 
where $s$ and $d$ span, in principle,  the whole Hilbert space, while on the other hand, for $s$, we use the Monte Carlo sampling so as to avoid the full use of the Hilbert space.
For $d$, we also restrict the whole Hilbert space by introducing a 
truncation scheme due to the overlap $(d |\psi)$. 
For a certain $\epsilon$, we can
truncate the coupled-cluster wave function as
\begin{equation}
 |\psi) \approx \sum_{|\psi(d)|^2>\epsilon}\psi(d)|d),
 \label{trunc_wf}
\end{equation}
where $\psi(d)=(d |\psi)$ is an amplitude of $|\psi)$ in the basis $d$.
 As the coupled-cluster wave function  $|\psi)$ is an expansion around the optimized deformed mean-field wave function $|0)$, we can naturally expect that this truncation scheme works well as we will see in Sec. III C.
Thus, we can also avoid the difficulty with the handling of the full Hilbert space for $d$.

\section{Benchmark test}

In this section, we want to demonstrate the feasibility  of our new method by taking the three-level Lipkin model as a testing ground.

\subsection{Three-level Lipkin model}

First, we define the three-level Lipkin model with the levels $i=0,1,2$,  each having a degeneracy of $N$. We consider $N$ spinless Fermions. The single particle energies, $\varepsilon_i$, are 
$\varepsilon_0=0$ and $\varepsilon_1=\varepsilon_2=1$. 
The creation and annihilation operators of the $k$-th particle on the $i$-th level are $c_{ik}^\dag$ and  $c_{ik}$, which we call, as before, spherical operators. 
The Hamiltonian is defined on this basis as 
\begin{equation}
H=\varepsilon(n_1+n_2) - \frac{v}{2}(J_{10}^2+J_{01}^2+J_{20}^2+J_{02}^2),
\end{equation}
where $\varepsilon=1$ and  $v$ is the strength of the two-body interaction. The
$J$ operators are defined by $ J_{pq}=\sum_i c_{p,i}^\dag c_{q,i}$.
The system is specified by the dimensionless parameter $\chi=v(N-1)/\varepsilon$. The spectra show two phases: One has a vibrational pattern around $\chi\sim 0$ and the other a rotational pattern for $\chi>2$. These two phases continuously change from one into the other when considering a finite number $N$.

The HF approximation is given by the mean-field state $ |\phi)=a^{\dag}_{0,1} \cdots a^{\dag}_{0,N}|-\rangle$
with the deformed operators, $a^{\dag}_{i,m}$ ($i=0 \sim 2$, $m=1 \sim N$), which are canonically related to the spherical ones with coefficients $D_{ij}$'s  as
$
 a^{\dag}_{i,m}=\sum _{j}D_{ij} c^{\dag}_{j,m}.
$
The HF energy can be shown to be $E_{HF}=0$ for $\chi<1$ and 
$E_{HF}=\frac{N\varepsilon}{4}(2-\chi-\frac{1}{\chi})$ for $\chi>1$  \cite{HB00}.

For the degenerate two upper single-particle energies, there is a symmetry concerning the exchange of 1 and 2 levels. This
 can be embodied by the symmetry operator (identifiable with a rotational operator) 
$L = i(J_{21}-J_{12})$
which commutes with the Hamiltonian, $[H,L]=0$. 
Thereby, all wave functions have the quantum number $L_0=0,\pm 1,\pm 2, \cdots$.
Except for $L_0=0$, the eigenstates with $L_0$ are doubly degenerate. 
With this symmetry operator, we can construct the projection operator 
 as 
\begin{equation}
P^{L_0}=\frac{1}{2\pi}\int_0^{2\pi} d\theta e^{i( L-L_0)\theta},
\label{def_L_proj}
\end{equation}
which projects out the $L_0$ component from any wave function $ |\psi) $ that is, 
$
 |\psi, L_0 ) = P^{L_0}|\psi)
$
where $L_0$ and $|\psi, L_0 )$ are eigenvalue and eigenstate of $L$, respectively, 
that is, $ L|\psi, L_0) = L_0|\psi, L_0 )$.
Note that $L$ is defined by the $c$ and $c^\dag$ operators
and the projection should be carried out using the spherical operators.

\subsection{Symmetry projection of CC wave function}

The CCSD wave function  $ |\psi )$ is given by 
$  |\psi ) =e^{Z}| 0)  $
where $|0)$ is the deformed HF state, and  $Z$ is organized into a sum of one-body and two-body operators
\begin{equation}
  Z=x_1 K_{10}+x_2 K_{20}+y_{11} K_{10}^2+y_{22} K_{20}^2+y_{12} K_{10}K_{20},
\end{equation}
where the $K$ operators are defined as $ K_{pq}=\sum_i a_{p,i}^\dag a_{q,i}$.
In applying the CCSD procedure described in Sec. II A,
we take the components $|1,0)$, $|0,1)$, $|1,1)$, $|2,0)$ and $|0,2)$ in the deformed basis, defined by 
$|k_1, k_2)=\mathcal{N}_{k_1,k_2}(K_{20})^{k_2}(K_{10})^{k_1}|0)$,
with  normalization factor $\mathcal{N}_{k_1,k_2}$.
With this, the CCSD equation can be solved easily. We can also carry out the 
CCD calculations by changing the mean-field.

\begin{figure}[h]
\includegraphics[width=7cm]{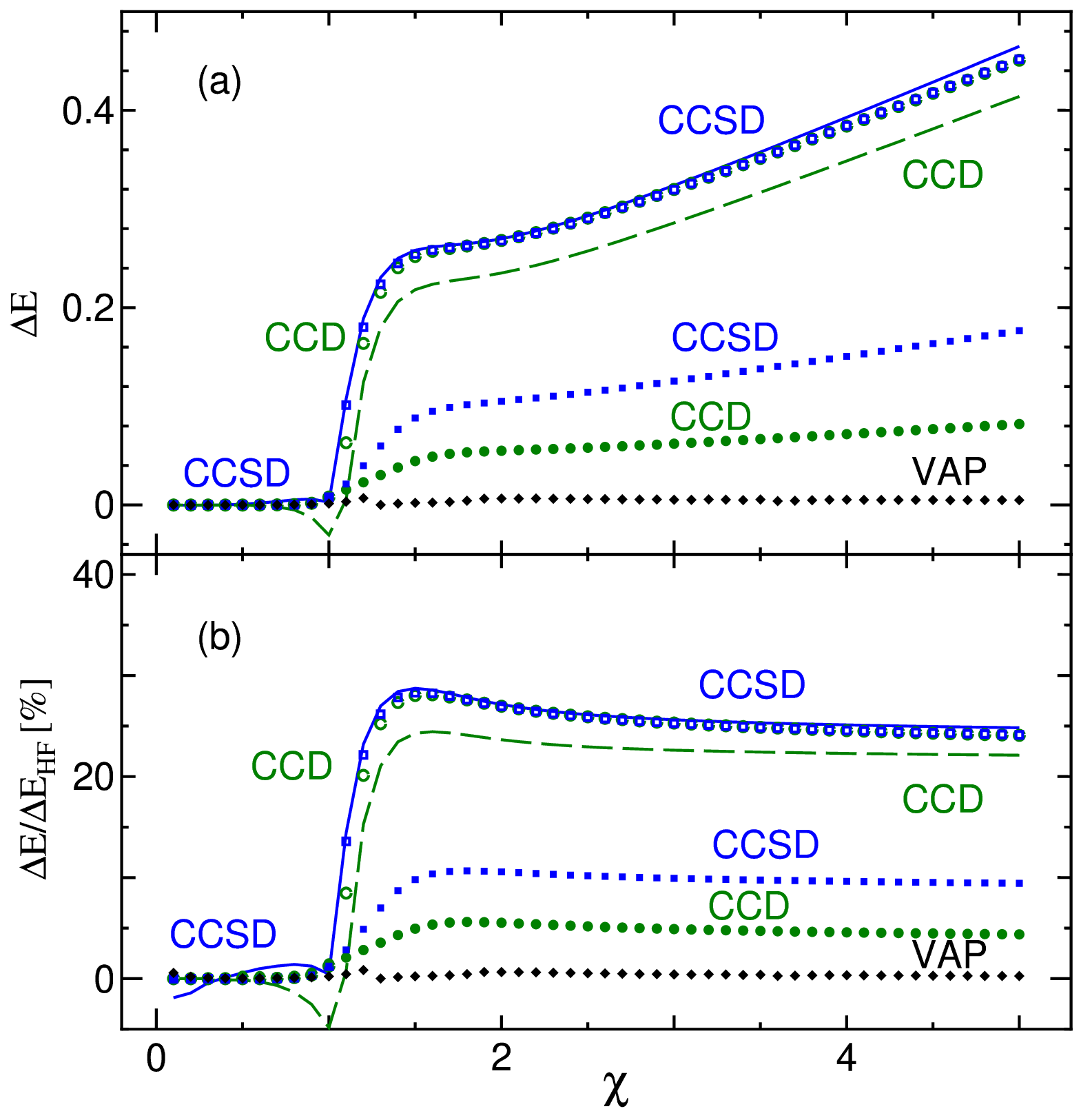}
\caption{
Energy differences $\Delta E$ (a) and ratios between $\Delta E$  and  $\Delta E_{HF}$ (b) are shown as a function of $\chi$ with $N=20$ ; 
(1) Blue (Green dashed) line for CC energy with the CCSD (CCD) wave function 
(2) Blue open squares (Green open dots) for RR energy with the CCSD (CCD) wave function.
(3)  Blue filled squares (Green filled dots) for projected energy with the CCSD (CCD) wave function
(4)  Black diamonds for VAP energy.
}\label{CCSD}
\end{figure}

Fig. 1(a) shows the CC energy differences $\Delta E= E-E_{exact}$ where this is relative to the exact ground state energy $E_{exact}$ as a function of $\chi$ for  $N = 20$.
In Fig. 1(b), as a measure of improvement, we plot the ratio between $\Delta E$ and  $\Delta E_{HF}=E_{HF}-E_{exact}$ which is the correlation energy.
It shows that the coupled-cluster wave function contains a considerable amount of more correlations. Moreover, the CC energy of CCD  is lower than that of CCSD.
As the coupled-cluster method does not, however, satisfy the Raleigh-Ritz variational principle, we calculate the RR energies in Eq.(\ref{var_energy}) with the same CCSD and CCD wave functions.
CCSD and CCD give almost the same RR energies.
The CCD wave function may seem not to be so good.  
It shows, however, a distinct aspect if we apply the symmetry projection. 
In Fig. 1, projected ground state energies with CCSD and CCD relative to the exact ones $\Delta E$ and $\Delta E/ \Delta E_{HF}$ are plotted.
The projected energy of the ground state is vastly improved over the CC energy, especially for the CCD, where about 95\% of correlation energy is taken into account. Note that VAP calculations can also be straightforwardly carried out in this simple model, and the calculated 
$\Delta E$ and  $\Delta E / \Delta E_{HF}$ are also plotted as a reference result. The VAP energy with  $\sim 10^{-4} - 10^{-5}$ 
 percentage error is almost perfect. We will, however, not further dwell on the VAP procedure, since it seems to be numerically intractable for realistic cases. On the other hand very recently the analog to the VAP calculation has been carried out for the pairing Hamiltonian in \cite{BD21}. We also will come back to the VAP approach in a follow-up paper in the near future.
\begin{figure}[h]
\includegraphics[width=7.2cm]{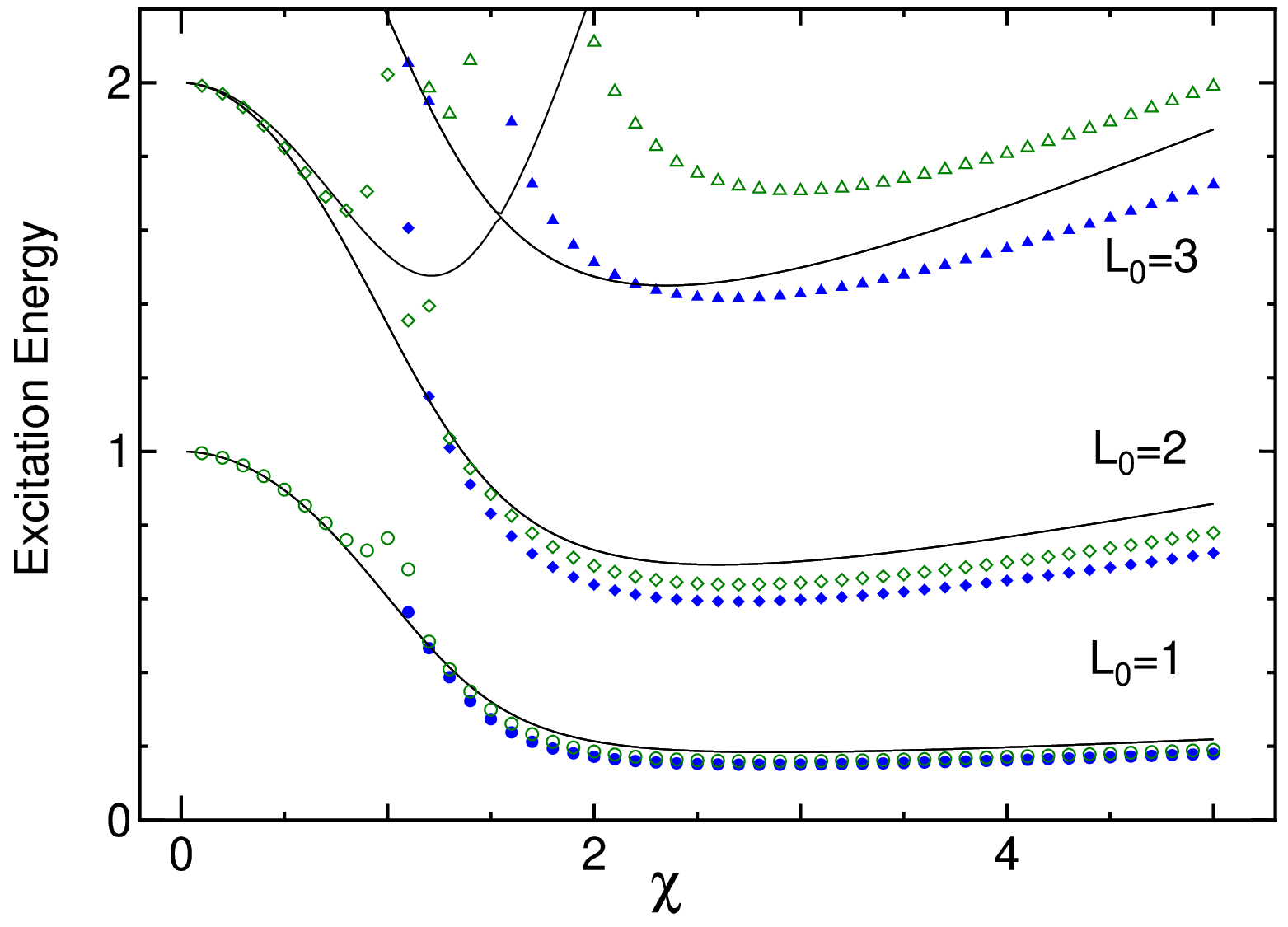}
\caption{
Excitation energies are shown as a function of $\chi$ with $N=20$;
(1) Black lines for exact energies ($L_0=1 \sim 3$)  (2) Blue filled (Green open) symbols for the projected energies ($L_0=1 \sim 3$) with the CCSD (CCD) wave function.
}\label{projectedCCSD}
\end{figure} 

In Fig. 2, the exact excitation energies for $L_0=1 \sim 3$ are plotted as a function of $\chi$ for  $N = 20$.
For $\chi\sim0$, spectra show a vibrational feature, while $\chi$ around 5, 
the spectra show a deformed band. For $1<\chi<2$, a crossover between these two phases occurs. 
We plot the projected excitation energies of Eq.(\ref{projected_energy}) with the CCSD and CCD wave functions, using the symmetry projection in Eq.(\ref{def_L_proj}) to show what results in the PACC procedure. The results show reasonably good excitation energies except for the crossover region.

For $0<\chi<1$, CCD and CCSD wave functions are the same by construction
and coupled-cluster method gives a pure spherical mean-field. 
Thereby, in the projected calculations with CCD, 
we added a slight deformation to the pure spherical mean-field and 
succeeded in reproducing the excitation energies.
Thus, the symmetry projection for the coupled-cluster wave function is quite
promising. The remainder of the problem is how to calculate such a symmetry projection
in realistic cases. Next, we introduce the Monte Carlo method and mixed representation for the projected overlap as in Eq.(\ref{sPLpsi}). 

\subsection{ Tests of Monte Carlo procedure}

The symmetry projection in Eq.(\ref{def_proj_mat}) requires heavy computations in realistic applications. So we introduce the Monte Carlo method as described in Sec. II C.
The spherical basis $s$ is sampled by the MCMC, which stochastically generates the distribution obeyed to $\rho_L(s)$ in Eq.(\ref{rho}).
To perform the MCMC, we take the basis states specified by $s$  as a random walker,  and we move this $s$-basis to another nearby $s'$-basis
under the control of the Metropolis-Hasting (MH) algorithm, keeping the detailed balance as in \cite{VMC12,VMC18}.

As a benchmark test of the MC calculation, we take the projected energy in Eq.(\ref{projected_energy}) with the CCD wave function, whose value for $L_0=0$  is  -17.789. 
We investigate the same projected energy by the Monte Carlo method.  
As the MC calculations have statistical errors, we examine the convergence by taking several sampling numbers $N_0=$ 30, 100, 300, 1000, 3000, 10000, 30000. For each $N_0$, we carry out 100 sets with different random numbers. 
In Fig.3, the MC energy in Eq.(\ref{mc_energy}) for each calculation is displayed. 
The average energy and its average variance over 100 sets are also plotted.
Thus, the statistical errors of the Monte Carlo procedure are well-controlled. 
The sampling number is, in general, relatively constant for various quantum systems with a larger Hilbert space.
\begin{figure}[h]
\includegraphics[width=7.5cm]{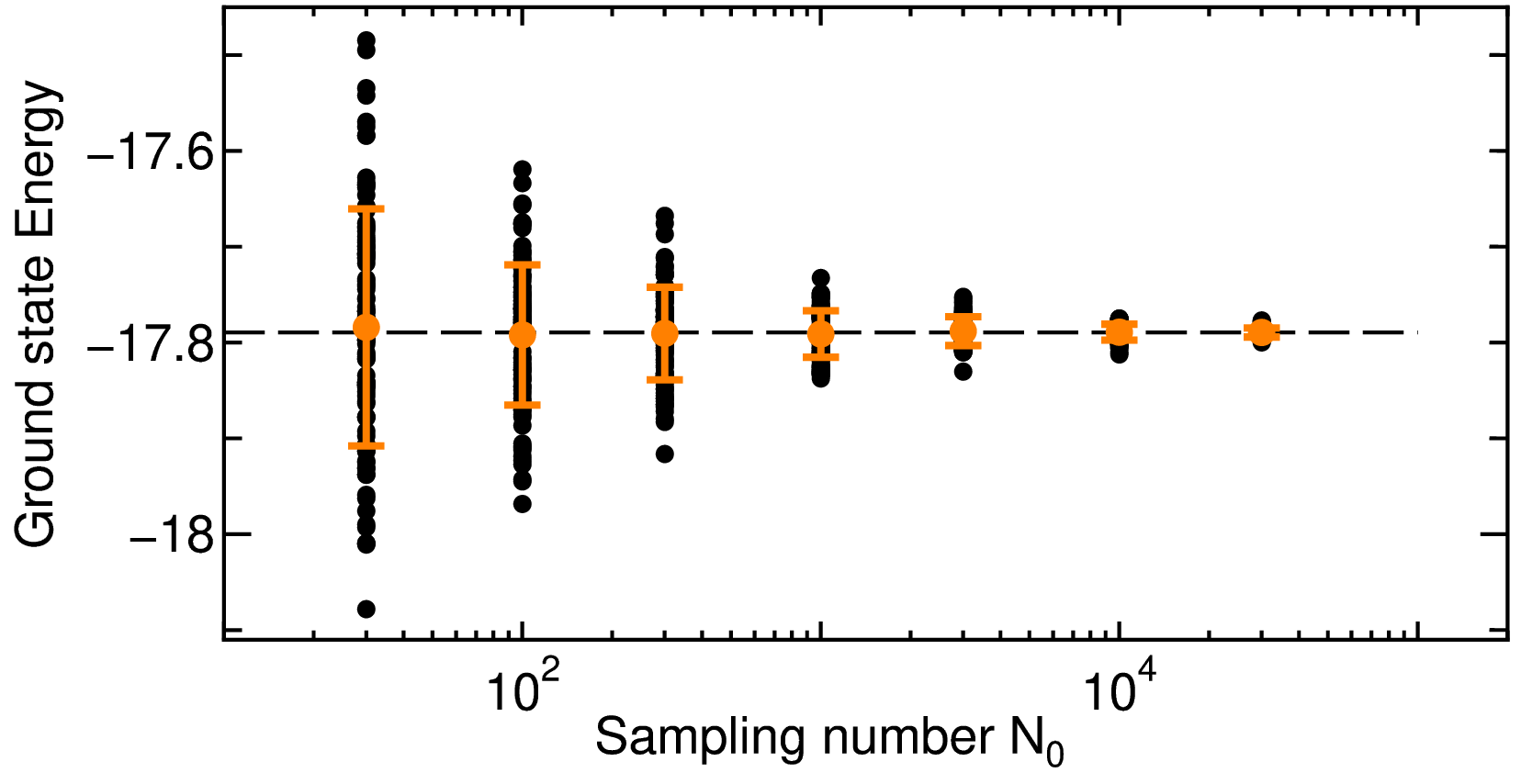}
\caption{
MC energies with $L_0=0$ as a function of Monte Carlo sampling number $N_0$.
For each  $N_0$, 100 sets of MC energies with different random numbers are plotted. Orange-filled circle and error bars stand for average energy and its one-standard deviations, respectively.
}\label{error_analysis}
\end{figure} 

Finally, we discuss the remaining problem: the computation of the projected overlaps $\langle s |P^{L_0}|\psi)$.
For the MC sampling in Eq.(\ref{localenergy}), the spherical basis is better
than the deformed one due to the advantage of the sparsity in $h_{s,s'}$. 
However, the projected overlap includes the exponential-type operators, of which exact projections become intractable for larger systems in general. Therefore we decompose the computation of the projected overlap as in Eq.(\ref{sPLpsi}), which allows us the following natural truncation of the deformed complete set.
 As the $|d)$ states are the ground state $|0)$ and all excited states thereof, the overlaps  $\langle s|P^{L_0}|\psi)$ for highly excited states $|d)$ are expected to be negligible, which then can be truncated by the value of $(d|\psi)$.
As the CCD wave function $|\psi)$ has no dependence of $\theta$ in Eq.(\ref{def_L_proj}) and is expressed by the deformed operators,
the numerical evaluation of  the values of $(d|\psi)$ is simple.
Moreover, contributions of higher excited states $|d)$ are, in general,  expected to be smaller. Therefore, the truncation of $d$ by $(d|\psi)$ is feasible.  
  
In Fig.4, we present a benchmark test of such a truncation scheme for the states with $L_0=0 \sim 3$. 
By setting a given threshold for the value $(d|\psi)$, we can truncate the summation for $d$ in Eq.(\ref{sPLpsi}). 
The MC energies in Eq.(\ref{mc_energy}) with these projected states are plotted as a function of the ratio of the truncated basis number to the one of the  whole  Hilbert space dimension. The parameters for the three-level Lipkin model are $\chi=5$ and $ N=20$.
The calculation at the ratio of 0.02 shows that the truncation scheme works quite well.
This ratio is expected to be increasingly smaller for larger systems.
Its investigation for various quantum systems will, however, be our task for the future.

\begin{figure}[h]
\includegraphics[width=7.5cm]{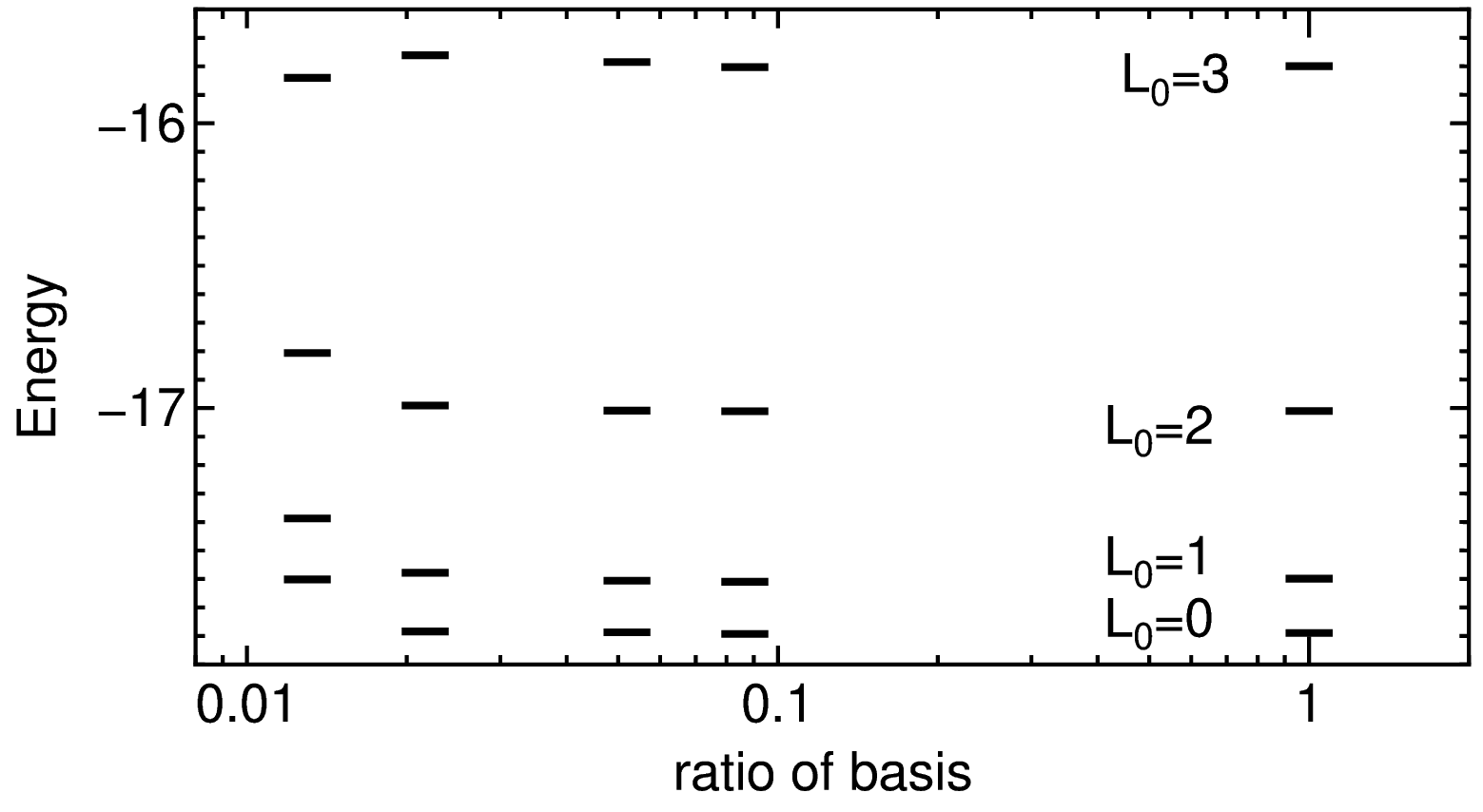}
\caption{
MC energies with $L_0=0\sim3$ as a function of basis truncation ratio.
At ratio 1, exact projected energies are also plotted.  The MC statistical errors
are invisible within the bars.
}\label{error_analysis}
\end{figure} 

\section{Summary}
This study combines the Raleigh-Ritz variational procedure, the coupled-cluster wave function, and symmetry projection through the Monte Carlo method. We start from the mean-field wave function and extend it by 
applying an exponential-type correlation factor $e^Z$  as within the coupled-cluster theory.
We use the coupled-cluster singles and doubles wave function
and its parameters are optimized by the coupled-cluster method
as in Eqs.(\ref{hbareq}-\ref{cceq}).
Next, we apply the symmetry projection to this wave function and
we directly evaluate the expectation value of the Hamiltonian with this projected wave function as in Eq. (\ref{projected_energy}).
Such an approach is, however, numerically intractable for practical applications.  Therefore we introduce the Monte Carlo method as in Eqs.(\ref{def_mc_energy}-\ref{mc_energy}).
We also introduce an efficient truncation scheme for the deformed set of states as in Eq.(\ref{trunc_wf}).

To evaluate the feasibility of this new method, we employ the three-level Lipkin model, which has spherical and deformed phases. Moreover, it has an exchange symmetry under degenerate single-particle energies of the upper two-levels. Therefore we can introduce the symmetry projection concerning 
this symmetry in Eq.(\ref{def_L_proj}).
By numerical calculations, we found that the symmetry projected CCD wave function can give a better ground state energy than the coupled-cluster method and gives reasonable spectra for low-lying states.
We showed that the Monte Carlo method and the truncation scheme we introduced also work quite well.

This new method can be applied to many quantum systems. For instance, the shell model calculation is being under investigation. 
The other direction of this method is to extend it to variation-after-projection, of which Monte Carlo realization is also under study. 

\section{Acknowledgements}
We acknowledge the financial support for a one-month stay in August/September, 2019 at LPMMC, Grenoble. The good working conditions have been appreciated.

\end{document}